\documentclass[prl,showpacs,floatfix,preprint]{revtex4}

\usepackage{epsfig,graphicx}

\def\fo{Fe$_3$O$_4$}
\def\gfo{$\gamma$-Fe$_2$O$_3$}
\def\Neel{N\'{e}el}

\begin{document}

\title{A method for measuring the {\Neel} relaxation time in a frozen ferrofluid}

\author{Ronald J. Tackett$^1$, Jagdish Thakur$^2$, Nathaniel Mosher$^1$, Emily Perkins-Harbin$^1$, Ronald E. Kumon$^1$, Lihua Wang$^3$, Corneliu Rablau$^1$, Prem P. Vaishnava$^1$\footnote{Corresponding author:  pvaishna@kettering.edu}}
\affiliation{$^1$Department of Physics, Kettering University, Flint, MI 48504}
\affiliation{$^2$Department of Physics and Astronomy, Wayne State University, Detroit, MI 48202}
\affiliation{$^3$Department of Chemistry and Biochemistry, Kettering University, Flint, MI 48504}

\date{\today}

\begin{abstract}
We report a novel method of determining the average {\Neel} relaxation time and its temperature dependence by calculating derivatives of the measured time dependence of temperature for a frozen ferrofluid exposed to an alternating magnetic field.  The ferrofluid, composed of dextran-coated {\fo} nanoparticles (diameter 13.7 nm $\pm$ 4.7 nm), was synthesized via wet chemical precipitation and characterized by x-ray diffraction and transmission electron microscopy.  An alternating magnetic field of constant amplitude ($H_{0}=20$ kA/m) driven at frequencies of 171 kHz, 232 kHz and 343 kHz was used to determine the temperature dependent magnetic energy absorption rate in the temperature range from 160 K to 210 K.  We found that the specific absorption rate of the ferrofluid decreased monotonically with temperature over this range at the given frequencies.  From these measured data, we determined the temperature dependence of the {\Neel} relaxation time and estimate a room-temperature magnetocrystalline anisotropy constant of 40 kJ/m$^3$, in agreement with previously published results.
\end{abstract}


\maketitle

\section{Introduction}
Colloidal suspensions of superparamagnetic iron-oxide nanoparticles, specifically those of magnetite ({\fo}) and maghemite ({\gfo}), have been extensively investigated for their potential applications such as cell separation, use as contrast agents for magnetic resonance imaging, targeted drug delivery and magnetic fluid hyperthermia (MFH) \cite{PankhurstJPhysD36, PankhurstJPhysD42}.  Among these applications, MFH has piqued the interest of researchers from various disciplines including biophysics, biomedicine and oncology due to its potential applications in various technologies for the treatment of cancer without the side effects inherent to radiation and chemotherapy-based methods \cite{DutzNanotech25}.  MFH involves the excitation of magnetic nanoparticles suspended in a fluid medium (a ferrofluid) using an oscillating magnetic field of the form $H(t)=H_{0}{\cos}(2{\pi}ft)$, where $H_{0}$ is the field amplitude (typically between 5 - 30 kA/m) and $f$ is the frequency (typically in the range from 150 kHz - 350 kHz).  For a single-domain nanoparticle, hysteresis losses are absent and energy absorption from the field can occur via two excitation mechanisms:  {\Neel} and Brownian relaxation.  This absorbed magnetic energy is eventually transformed to thermal energy and the temperature of the ferrofluid rises.  {\Neel} relaxation \cite{BrownPhysRev130} involves the alignment of the nanoparticle moment with the external field within a fixed, non-rotating nanoparticle.  As the excitation occurs against the anisotropy energy barrier of the particle the relaxation time depends strongly on the nanoparticle's magnetic volume, $V_{m}$, magnetocrystalline anisotropy constant, $K$, temperature, $T$ and characteristic relaxation time, ${\tau}_{0}$.  While there are a few models to describe this relaxation, the most commonly used has the form \cite{BrownPhysRev130}
\begin{equation}
\tau_{N}=\frac{\sqrt{{\pi}}}{2}{\tau}_{0}{\exp}\left(\frac{KV_{m}}{k_{B}T}\right)\sqrt{\frac{KV_{m}}{k_{b}T}}.
\label{EQ:NeelRelax}
\end{equation}
Brownian relaxation \cite{DebyePolarMolecules} involves the alignment of the particle's moment with the external field via the physical rotation of a fixed-moment nanoparticle within the carrier fluid.  This alignment is affected by the hydrodynamic properties of the carrier fluid and nanoparticles, it is described by \cite{FrenkelBook}
\begin{equation}
{\tau}_{B}=\frac{3{\eta}V_{H}}{k_{B}T},
\label{EQ:BrownRelax}
\end{equation}
where ${\eta}$ is the viscosity of the carrier fluid and $V_{H}$ is the hydrodynamic volume of the composite particle which includes any surfactant layer used to give colloidal stability to the ferrofluid.  When both mechanisms are active these processes occur in parallel and the effective relaxation time which describes the energy transfer rate is given by
\begin{equation}
{\tau}=\frac{{\tau}_{N}{\tau}_{B}}{{\tau}_{N}+{\tau}_{B}}.
\label{EQ:EffectiveRelax}
\end{equation}
From this relation, it is evident that the shorter relaxation time dominates the overall dissipative characteristics and thus determines the heating rate of the sample.

The {\Neel} mechanism plays a dominant role in the relaxational characteristics of the nanoparticles particularly when they are embedded in a cancerous tissue undergoing MFH treatment, since in the tissue's environment, due to immobilization of the particles, the Brownian mechanism is highly damped.  Several studies have shown that if the magnetic nanoparticles are internalized by the cancer cells, they are either locked in the cell plasma or adhere to the cell walls \cite{TellingACSnano9}.  Consequently, the role of Brownian relaxation is insignificant for nanoparticles used in the hyperthermia treatment of cancer when the ferrofluid is applied directly to the tumor tissues via injection.  In such situations, {\Neel} relaxation produces the local heating of the cancerous mass; however, one must be careful about comparing experiments done in the laboratory with therapies that will be performed under real-life conditions.

The experimental studies of the heating rates of ferrofluids in the presence of alternating magnetic fields can be broadly placed into one of two categories:  those focusing on varying only the intrinsic physical and magnetic parameters of the nanoparticles \cite{DewhirstCancerRes60,TackettJMMM320} and those that investigate the influence of varying only the parameters of the applied field (i.e. $H_{0}$ and $f$) \cite{WeitschiesJPhysCondMat18}.  Among these two broad classifications, studies of how the intrinsic parameters of nanoparticles influence magnetic heating still remain an active area of research.  However, as the effects of the intrinsic parameters are reflected through the relaxation processes ({\Neel} and Brownian) which act in parallel, it is difficult to understand the contribution of each process on the heating characteristics of the ferrofluid.

Knowing the significance of the {\Neel} mechanism in hyperthermia, many studies have been performed to estimate this parameter using ac and dc magnetic susceptibility measurements \cite{FanninJPhysD27,KolitzJMMM149}; however, these methods often require large volumes of sample, accurate determinations of sample volume and mass and costly equipment.  In this paper, we present a new method to determine this parameter using a simple induction heating system.  This method requires only measurements of the sample heating rate under the influence of an oscillating magnetic field of constant amplitude at two different frequencies.  Additional measurements of sample mass, volume and magnetic anisotropy are not needed.  The goal of this investigation is to determine the temperature dependence of the {\Neel} relaxation time (${\tau}_{N}$) when the Brownian mechanism is completely quenched.  This suppression of the Brownian relaxation is achieved by performing the experiments in the temperature range from 160 K to 210 K where the carrier fluid (DI water) is completely frozen, thus locking the nanoparticles in place.

Experimentally, the heating rate in MFH is expressed in terms of the specific absorption rate ($SAR$) which is defined to be the power absorbed per unit mass of the nanoparticles in the ferrofluid.  This can be calculated using the thermodynamic relation for rate of heat energy absorbed by the sample as \cite{BabincovaJMMM225}
\begin{equation}
SAR_{heating}=\frac{M_{sample}}{m_{Fe_{3}O_{4}}}C(T)\frac{{\Delta}T_{heating}}{{\Delta}t},
\label{EQ:SARheating}
\end{equation}
where $M_{sample}$ is the overall mass of the sample, $m_{Fe_{3}O_{4}}$ is the mass of the {\fo} nanoparticles, $C(T)$ is the temperature dependent specific heat of the carrier fluid and ${\Delta}T_{heating}/{\Delta}t$ is the time rate of change of the sample's temperature as it absorbs energy from the applied magnetic field.  When the sample begins to warm under the influence of the applied magnetic field, it is important to note that heat exchange with the environment is also occurring due to the temperature differential between the sample and its surroundings.  For accurate determination of $SAR$, it is important to estimate this heat exchange between the sample and the environment due to convective, conductive and radiative processes.  While this process is quite hard to do from a calculational standpoint, we can experimentally estimate the collective effects of heat exchange via these processes by measuring the heating of the sample placed into the experimental apparatus while keeping the external field turned off.  From this data, we determine the specific power gain (SPG) as heat is transferred from the environment to the sample as \cite{NemalaJAP116}
\begin{equation}
SPG_{warming}=\frac{M_{sample}}{m_{Fe_{3}O_{4}}}C(T)\frac{{\Delta}T_{warming}}{{\Delta}t},
\label{EQ:SPGwarming}
\end{equation}
where ${\Delta}T_{warming}/{\Delta}t$ is the warming rate of the sample due only to heat exchange between the sample and the environment.  Once this SPG is determined, we can correct the measured $SAR$ for the sample as a function of temperature \cite{NemalaJAP116}
\begin{equation}
SAR=SAR_{heating}-SPG_{warming}
\label{EQ:SARcorrected}
\end{equation}
Due to the suppression of the Brownian mechanism, the values of $SAR$ obtained over this temperature interval are dependent only upon {\Neel} relaxation.  This relaxation time can be determined using a theoretical model based on a direct relationship between the power density dissipated by the nanoparticles and the out-of-phase, dissipative component of the ferrofluid's susceptibility \cite{RosensweigJMMM252}
\begin{equation}
P={\pi}{\mu}_{0}H_{0}^{2}{\chi}_{0}f\frac{2{\pi}f{\tau}_{N}}{1+(2{\pi}f{\tau}_{N})^{2}},
\label{EQ:PowerDensity}
\end{equation}
where ${\chi}_{0}$ is the equilibrium susceptibility and ${\mu}_{0}$ is the permeability constant.  In this model, the energy dissipation by the nanoparticles, which is related to the out-of-phase component of the magnetic susceptibility, is expressed in terms of the equilibrium susceptibility and the {\Neel} relaxation time ${\tau}_{N}$.  Measuring the $SAR$ at two different frequencies $f_{1}$ and $f_{2}$ while keeping the field amplitude ($H_{0}$) fixed, we find ${\tau}_{N}$ from Equation \ref{EQ:PowerDensity} as
\begin{equation}
{\tau}_{N}=\frac{1}{2{\pi}f_{1}f_{2}}\sqrt{\frac{f_{1}^{2}-{\alpha}f_{2}^{2}}{{\alpha}-1}},
\label{EQ:TauNeelAlpha}
\end{equation}
where
\begin{equation}
\alpha=\frac{SAR_{f_{1}}}{SAR_{f_{2}}}=\frac{\left(\frac{{\Delta}T_{heating}}{{\Delta}t}-\frac{{\Delta}T_{warming}}{{\Delta}t}\right)_{f_{1}}}{\left(\frac{{\Delta}T_{heating}}{{\Delta}t}-\frac{{\Delta}T_{warming}}{{\Delta}t}\right)_{f_{1}}}.
\label{EQ:alpha}
\end{equation}
It is interesting to note that ${\tau}_{N}$ is determined only from the values of $f_{1}$, $f_{2}$ and the sample heating rates (corrected for heat exchange with the environment) and do not require measurement of the sample mass, nanoparticle mass, or specific heat values.

\section{Experimental Details}
Iron oxide nanoparticles were synthesized using a standard co-precipitation technique in which an aqueous solution of FeCl$_{3}\cdot$6H$_{2}$O and FeCl$_{2}\cdot$4H$_{2}$O were mixed in a 2:1 molar ratio and {\fo} nanoparticles were precipitated by the drop-wise addition of 1M NH$_{4}$OH.  During precipitation, N$_{2}$ gas was bubbled through the solution to protect against oxidation of the Fe$^{2+}$ ions into Fe$^{3+}$ ions.  The precipitate was separated from the solution by a strong magnet, washed with DI water and re-suspended in a metastable 0.5 M NaOH solution. In order to suspend the precipitated nanoparticles in a carrier solution (DI water) they were coated in dextran by the drop-by-drop addition of the metastable solution of {\fo} to a solution of 15-20 kDa dextran (MP Biomedicals) in 0.5M NaOH while simultaneously probe sonicating.  The product was rinsed and resulted in a water-based suspension of {\fo} nanoparticles with a concentration of 20 mg of {\fo} per mL of solution.  A portion of the sample was lypholized and characterized via x-ray diffraction (Rigaku MiniFlex 600) and transmission electron microscopy (JEOL HR TEM 2010 operating at 200 keV).

For calorimetric measurements in the 160 K to 210 K range, the ferrofluid sample was first cooled to 77 K via the immersion of seald vials in liquid nitrogen.  The sample was then allowed to warm under ambient conditions to 160 K at which time a ~20 kA/m (250 Oe) ac magnetic field was applied using an Ambrell EasyHeat 2.4 kW induction heating system with a water-cooled, 8-turn, 2-cm-diameter coil.  Temperature versus time data were collected in the 160 K to 210 K region using an Optocon FOTEMP1-H fiber optic temperature monitoring system equipped with a TS5 optical temperature sensor with 0.1 K accuracy.  The sample vial was thermally insulated using cotton padding and foam rubber to minimize heat exchange with the environment; however, as thermal interaction with the environment cannot be completely suppressed, experiments were performed to determine the extent of the heat exchange with the surrounding environment and this estimation was accounted for in all measurements as discussed above.

\section{Results and Discussion}
\begin{figure} \smallskip \centering 
\epsfig{file=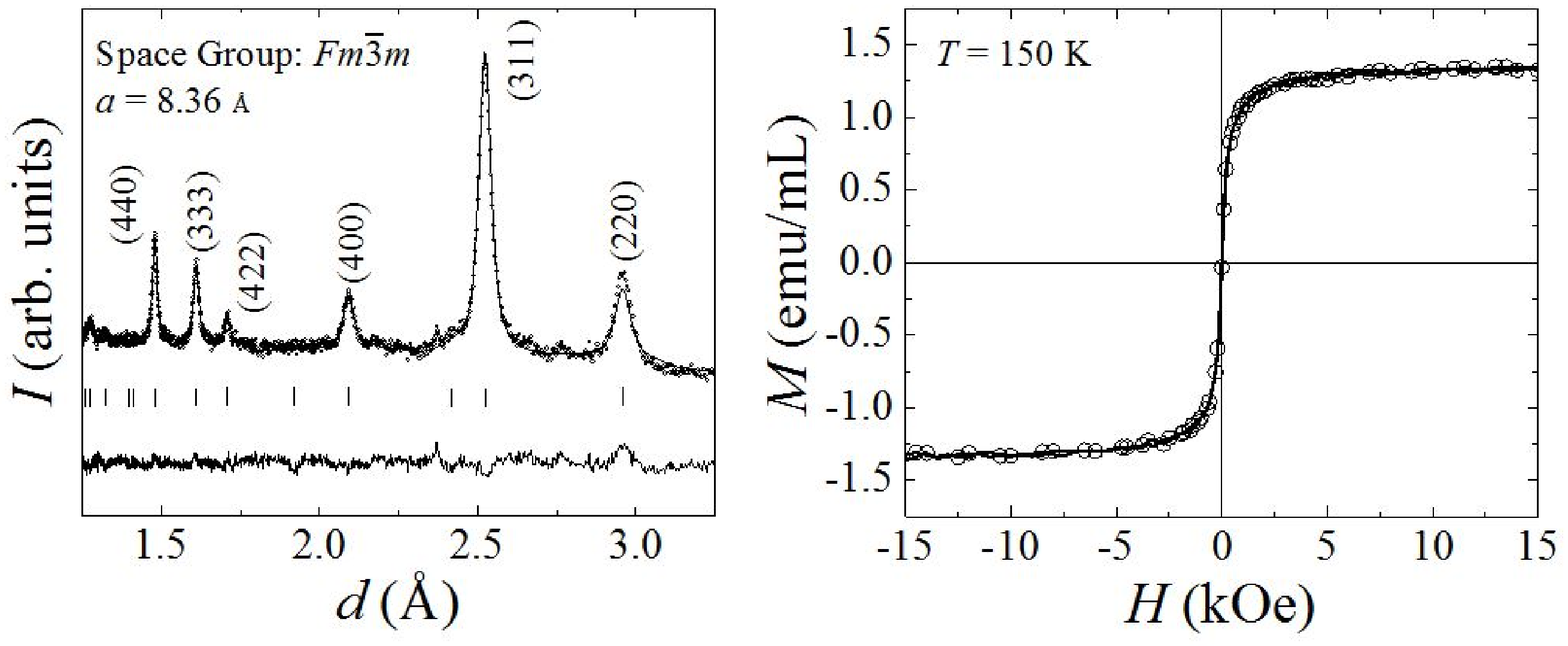, width=12cm} 
\caption{(Left Panel) A powder x-ray diffraction spectrum of a lypholized portion of the sample.  The open symbols correspond to the oberserved data and the solid line is a full-profile Le Bail fit to the data indicated the $Fm\bar{3}m$ cubic symmetry and 8.36 {\AA} lattice constant characteristic of {\fo}.  The Miller indices of the more intense reflections are indicated and the bottom trace is the difference curve between the observed data and the fit.  (Right Panel) The magnetic field dependence of the sample's magnetization at 150 K showing that the nanoparticles are in the superparamagnetic state in the experimental temperature regime between 160 K and 210 K.}
\label{FIG:XRDmag} 
\end{figure}

The left panel of Figure \ref{FIG:XRDmag} shows the powder x-ray diffraction (XRD) spectrum taken from the lypholized dextran coated {\fo} nanoparticles.  The open symbols represent the observed counts recorded for different $d$-spacing values between 1.25 and 3.25 {\AA}.  The solid line is a full-profile Le Bail fit to the data, the vertical bars indicated the $d$-spacing positions of the Bragg reflections and the lower trace is the difference curve between the observed and calculated counts.  The fit confirms that the sample consists of a single nanocrystalline phase of {\fo} with cubic $Fm\bar{3}m$ symmetry and lattice constant $a$ = 8.36 {\AA}.  Using the full-width at half maximum of the (311) reflection in Scherrer's equation, an average nanoparticle diameter of 14 nm was estimated for this sample.  The right panel of Figure \ref{FIG:XRDmag} shows the magnetic field dependence of the ferrofluid's magnetization measured at 150 K which exhibits no hysteresis thus confirming the superparamagnetic nature of the nanoparticles within the experimental temperature range between 160 K and 210 K.

\begin{figure} \smallskip \centering 
\epsfig{file=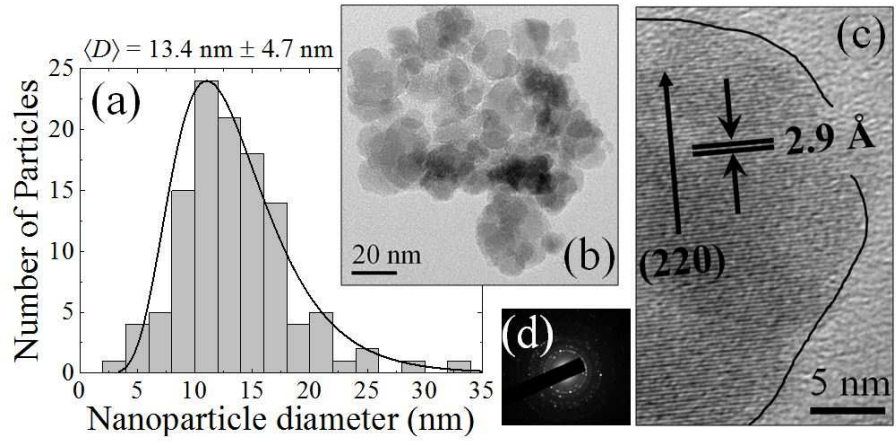, width=12cm} 
\caption{(a) A histogram of the particle size distribution determined using the micrograph shown in (b) and other similar micrographs taken at the same magnification. (c) A high-resolution TEM micrograph of a portion of a single nanoparticle with an outline of the particle provided for clarity.  The spacing of the observed lattice planes was found to be 2.9 {\AA} indicative of the (220) set of planes.  (d) A selected area electron diffraction (SAED) pattern which was used to confirm the cubic nature of the sample.}
\label{FIG:TEM} 
\end{figure}

Transmission electron microscopy (TEM) was performed to determine the mean diameter and size distribution of the nanoparticles in addition to confirming the crystallographic information determined via XRD.  Figure \ref{FIG:TEM}(a) shows a histogram of particle sizes determined from the bright-field TEM micrograph shown in Figure \ref{FIG:TEM}(b) and similar images at the same magnification.  The histogram shows a log-normal particle size distribution having mean diameter ${\langle}D{\rangle}$ = 13.4 nm and standard deviation from the mean of ${\sigma}_{D}$ = 4.7 nm.  This number is in agreement with the 14 nm average diameter determined using Scherrer's equation in conjunction with the collected XRD spectrum.  Figure \ref{FIG:TEM}(c) shows a high-resolution TEM micrograph of a portion of a single nanoparticle.  The visible lattice planes in the image show a spacing of 2.9 {\AA} marking them as the (220) set of planes in {\fo}.  Lastly, Figure \ref{FIG:TEM}(d) shows a selected area electron diffraction (SAED) pattern which was used to to confirm the cubic symmetry determined using XRD.

\begin{figure} \smallskip \centering 
\epsfig{file=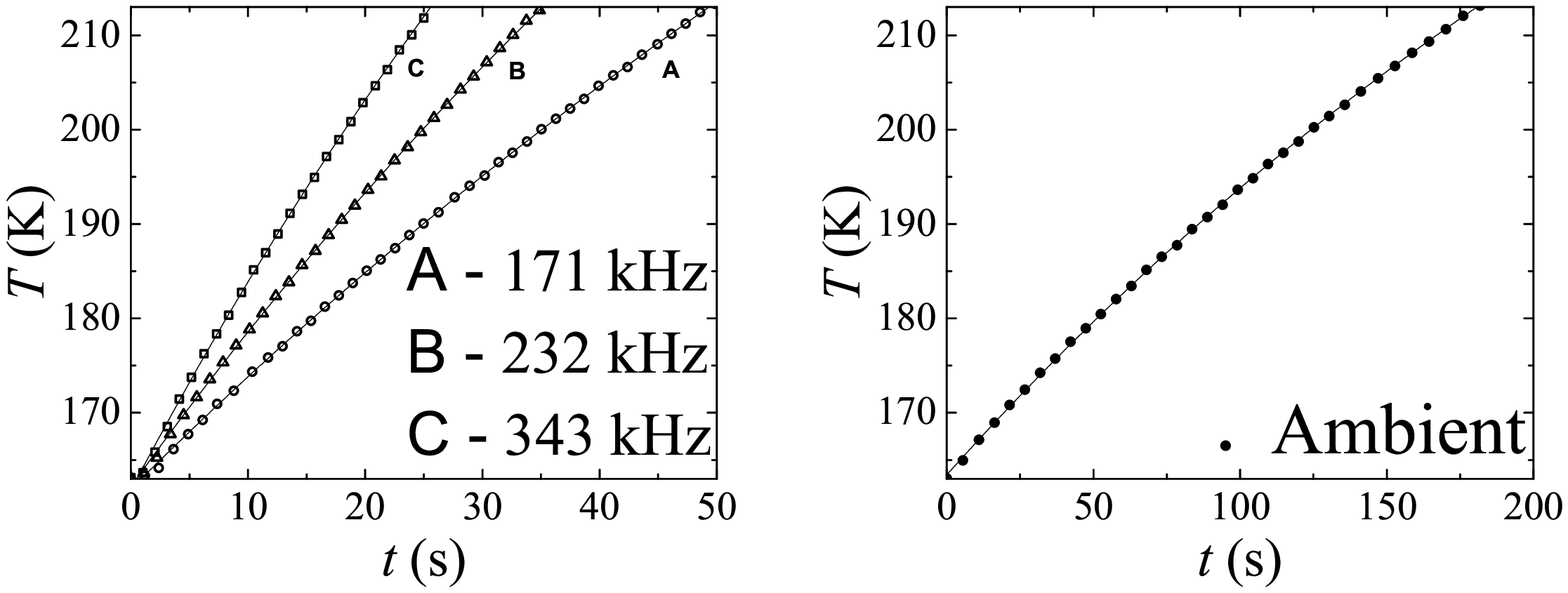, width=15cm} 
\caption{(Left Panel) Temperature versus time measurements made with the ferrofluid in an alternating magnetic field of amplitude $H_{0}$ = 20 kA/m at frequencies of 171 kHz (A), 232 kHz (B) and 343 kHz (C).  (Right Panel) Temperature versus time measurements made with the ferrofluid in the field coil but in the absence of an applied field.}
\label{FIG:HeatingCurves} 
\end{figure}

\begin{figure} \smallskip \centering 
\epsfig{file=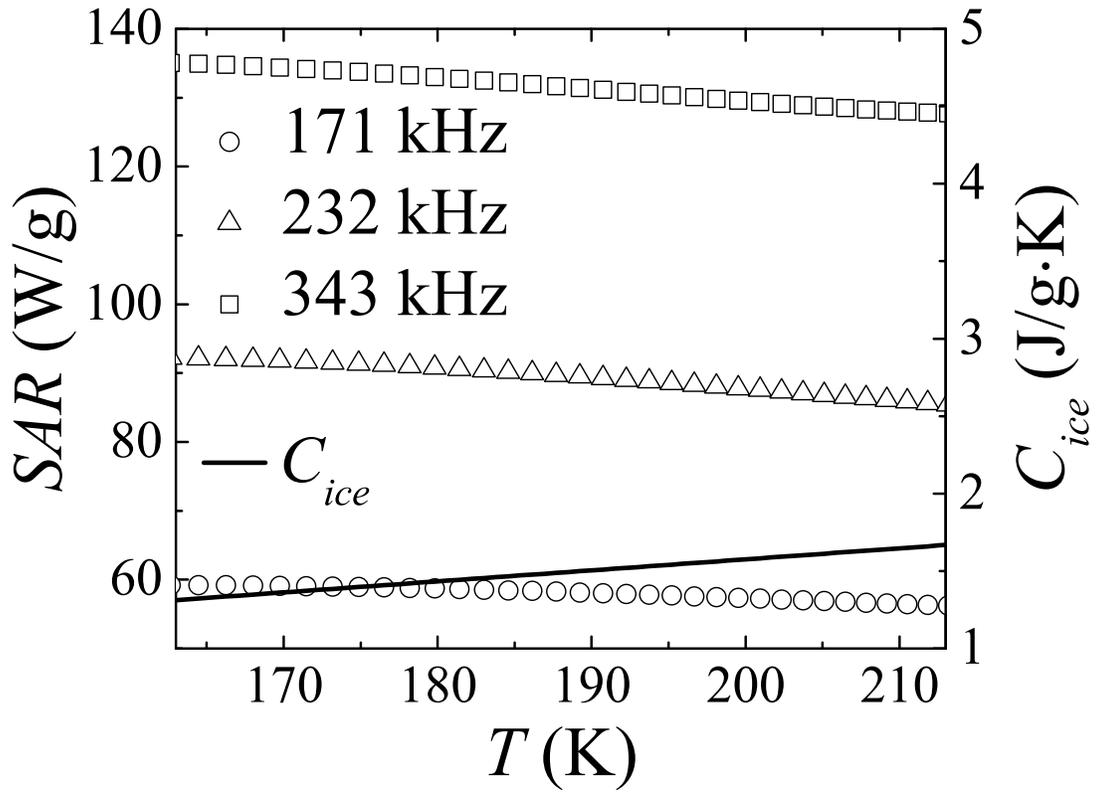, width=15cm} 
\caption{The temperature dependence of the specific absorption rate ($SAR$) of the ferrofluid measured at frequencies of 171 kHz (open circles), 232 kHz (open triangles) and 343 kHz (open squares).  The temperature dependence of the specific heat of ice used to calculate the specific absorption rate is plotted as a solid line}
\label{FIG:SARcurves} 
\end{figure}

The left panel of Figure \ref{FIG:HeatingCurves} shows the temperature versus time data collected while heating the ferrofluid using an alternating magnetic field of amplitude $H_{0}$ = 20 kA/m at 171 kHz (open circles), 232 kHz (open triangles) and 343 kHz (open squares).  All three curves were fit to a polynomial (solid lines) and differentiated in order to determine the slope, ${\Delta}T/{\Delta}t$.  To account for any heat exchange with the environment, a control experiment was performed in which the ferrofluid was placed inside the field coil and allowed to warm in the absence of the magnetic field.  This data is plotted in the right panel of Figure \ref{FIG:HeatingCurves} and will be referred to as the ambient curve.  From this data, we were able to calculate the specific absorption rate ($SAR_{heating}$) for each frequency and the specific power gain ($SPG_{warming}$) for the ambient experiment all as functions of temperature using the temperature-dependent specific heat of ice \cite{CRRELreport81} and Equations \ref{EQ:SARheating} and \ref{EQ:SPGwarming}, respectively.  From these calculations, we were able to determine the specific absorption rate ($SAR$) due only to energy absorbed from the alternating magnetic field using Equation \ref{EQ:SARcorrected}.  The results of these calculations are shown in Figure \ref{FIG:SARcurves} and yield average values of $SAR$ over the temperature interval of 57 W/g, 88 W/g and 130 W/g at 171 kHz, 232 kHz and 343 kHz, respectively.

\begin{figure} \smallskip \centering 
\epsfig{file=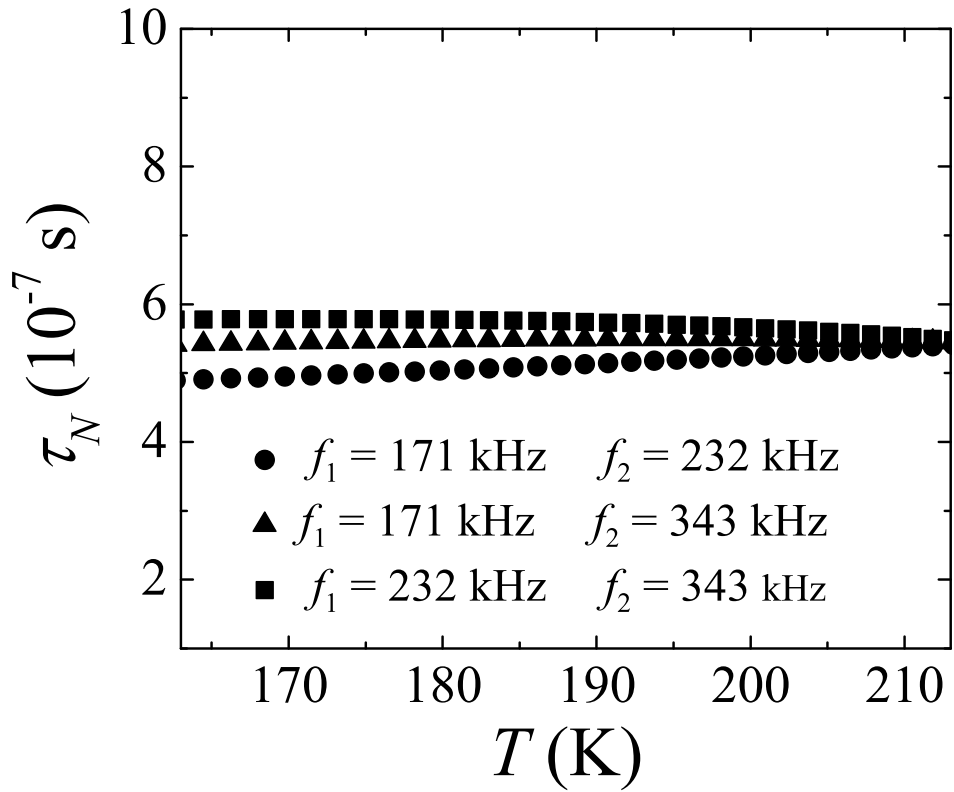, width=15cm} 
\caption{The temperature dependence of the {\Neel} relaxation time calculated using equations 8 and 9 for frequencies $f_{1}$ = 171 kHz and $f_{2}$ = 232 kHz (open circles), $f_{1}$ = 171 kHz and $f_{2}$ = 343 kHz (open triangles) and $f_{1}$ = 232 kHz and $f_{2}$ = 343 kHz (open stars).}
\label{FIG:TauCurves} 
\end{figure}

Additionally, using the temperature dependent slopes of the data presented in Figure \ref{FIG:HeatingCurves} found through differentiation of the polynomial fits, we are able to use Equations \ref{EQ:TauNeelAlpha} and \ref{EQ:alpha} to calculate the temperature dependence of the {\Neel} relaxation time over the temperature regime in question.  The results of this calculation are shown in Figure \ref{FIG:TauCurves} and show a value near 5 $\times$ 10$^{-7}$ s for all possible permutations of $f_{1}$ and $f_{2}$.  Using Equation \ref{EQ:NeelRelax} with this determined value of ${\tau}_{N}$, the experimentally determined particle diameter of 13.7 nm, and assuming a characteristic relaxation time of ${\tau}_{0}$ = 10$^{-9}$ s we find the room temperature magnetocrystalline anisotropy constant, $K$, for this sample to be 40 kJ/m$^{3}$, which agrees with the range of values previously reported in the literature \cite{GoyaJAP94, CaruntuJPhysD40}.

It is important to emphasize that for quantitative determination of ${\tau}_{N}$ using Equation \ref{EQ:TauNeelAlpha}, one needs the time derivatives of the temperature at two different frequencies and for the ambient (no applied field) case.  Using these values in Equation \ref{EQ:alpha}, we determine the temperature dependence of the {\Neel} relaxation time using Equation \ref{EQ:TauNeelAlpha}.  In our case, by taking data at three different frequencies, we were able to provide three different estimations of the temperature dependence of ${\tau}_{N}$ and found, as expected, that the value was roughly the same for each pair of frequencies.  We see in Figure \ref{FIG:TauCurves} a convergence of the values of ${\tau}_{N}$ as we approach 210 K.  In addition, it is important to note that the magnitude of the {\Neel} relaxation time is on the order of 10$^{-7}$ s as expected for a system of non-interacting magnetic nanoparticles having characteristic time constant ${\tau}_{0}{\sim}10^{-9} - 10^{-13}$ s \cite{BrownPhysRev130,WernsdorferPRL78}.  The non-interacting nature of this ensemble was confirmed through calculation of the relative variation of the blocking temperature per frequency decade (${\phi}={\Delta}T/T{\log}_{10}f$) in which we found a value of ${\phi}=0.18$, an order of magnitude larger than that found in interacting samples exhibiting spin-glass-like transitions \cite{KodamaPRL77,TholenceSSComm49,DuttaJPhysD43,PisaneJMMM384}.  The details of this investigation will be published in a subsequent paper \cite{TackettTBA}. At the macroscopic level, ${\tau}_{N}$, as described by Equation \ref{EQ:NeelRelax}, references the relaxation of an individual nanoparticle of volume $V_{m}$ having magnetocrystalline anisotropy constant $K$ and characteristic relaxation time ${\tau}_{0}=\sqrt{{\pi}KV_{m}/4k_{B}T}$; however, in a polydisperse system these parameters may vary from particle to particle and consequently their {\Neel} relaxation times do vary.  In $SAR$ measurements, one measures the temperature rise of the ferrofluid due to the average thermal energy of its constituents (i.e. nanoparticles, surfactant and carrier fluid) in contrast to the relaxation time described by Equation \ref{EQ:NeelRelax} which refers to a single particle.  Thus, our measured ${\tau}_{N}$ represents an average {\Neel} relaxation time of the ensemble of particles--this is what is responsible for the measured values of $SAR$.  However, for a monodisperse system, Equations \ref{EQ:NeelRelax} and \ref{EQ:TauNeelAlpha} may be applied as what is true for any one particle is true for any other in the system.  The use of Equation \ref{EQ:NeelRelax}, however, requires an accurate determination of the particle volume, $V_{m}$, and the magnetocrystalline anisotropy constant, $K$.  On the other hand, the method outlined in this paper requires only the value of the applied field frequencies and the derivatives of the temperature versus time curves which can be accurately determined using simple to understand methodology.

\section{Conclusion}
We studied the temperature dependence of the specific absorption rate ($SAR$) of a frozen ferrofluid at different frequencies at a constant field amplitude.  From the determined $SAR$ values we were able to calculate the temperature dependence of the {\Neel} relaxation time using a novel and simple approach.  This method is useful for cases in which there are uncertainties related to the magnetocrystalline anisotropy constant, $K$, and nanoparticle size.  In our experimental temperature window between 160 K and 210 K, our measured $SAR$ decreases with temperature; however, this trend may not continue as the experimental window is widened to include temperatures above 210 K where other effects may begin to play a role in the relaxation characteristics of the nanoparticles.  We believe that this investigation offers a new and simpler method of determining an average {\Neel} relaxation time in a colloidal suspension of magnetic nanoparticles.

\section{Acknowledgements}
This work was supported by the NSF though DMR-1337615 which provided funding for the Miniflex 600 powder x-ray diffractometer.  RJT, REK, CR and PPV would like to thank Dr. James Zhang for his support through the Faculty Research Fellowship.  In addition, PPV would like to thank Dr. Zhang for his support through the Rodes Professorship.  Lastly, REK would like to thank the Society of Physics Students for the support.

\clearpage

\clearpage

\clearpage

\clearpage

\clearpage

\end{document}